\documentclass[runningheads,a4paper]{llncs}
\usepackage{makeidx}
\usepackage{amssymb}
\usepackage{amscd, amsfonts}
\usepackage{verbatim}
\usepackage{listings}
\usepackage{commath}
\usepackage{float}
\usepackage{tikz}
\usepackage{isabelle}
\usepackage{isabellesym}

\usepackage{lmodern}
\usepackage[T1]{fontenc}

\newcommand{\isasty}[1]{\textbf{\textit{#1}}}

\newcommand{\isato}{\isasymrightarrow}
\newcommand{\isaTo}{\isasymRightarrow}
\newcommand{\isalb}{\isacharbraceleft}
\newcommand{\isarb}{\isacharbraceright}
\newcommand{\isaset}[1]{\isalb #1\isarb}
\newcommand{\isain}{\isasymin}
\newcommand{\isaprod}{\isasymtimes}
\newcommand{\isaeval}{\isamath{^\backprime}}
\newcommand{\isaimage}{\isamath{^{\backprime\backprime}}}

\newcommand{\isapair}[1]{\isasymlangle #1\isasymrangle}
\newcommand{\isaiff}{\isasymlongleftrightarrow}
\newcommand{\isaimp}{\isasymlongrightarrow}
\newcommand{\isaImp}{\isasymLongrightarrow}
\newcommand{\isasub}[2]{\isamath{#1_#2}}
\newcommand{\isazero}[1]{\isasub{\mathbf{0}}{\isasty{#1}}}
\newcommand{\isaone}[1]{\isasub{\mathbf{1}}{\isasty{#1}}}
\newcommand{\isaplus}[1]{\isasub{+}{\isasty{#1}}}
\newcommand{\isaminus}[1]{\isasub{-}{\isasty{#1}}}
\newcommand{\isatimes}[1]{\isasub{*}{\isasty{#1}}}
\newcommand{\isaplusN}{\isasub{+}{\mathbb{N}}}
\newcommand{\isaplusR}{\isasub{+}{\mathbb{R}}}
\newcommand{\isaminusR}{\isasub{-}{\mathbb{R}}}
\newcommand{\isatimesR}{\isasub{*}{\mathbb{R}}}
\newcommand{\isale}[1]{\isasub{\le}{\isasty{#1}}}
\newcommand{\isageN}{\isasub{\ge}{\mathbb{N}}}
\newcommand{\isageR}{\isasub{\ge}{\mathbb{R}}}
\newcommand{\isaltR}{\isasub{<}{\mathbb{R}}}
\newcommand{\isagtR}{\isasub{>}{\mathbb{R}}}

\newcommand{\isaempty}{\isasymemptyset}
\newcommand{\isamor}{\isasymrightharpoonup}
\newcommand{\isazeroR}{\isamath{0_\mathbb{R}}}
\newcommand{\isaoneR}{\isamath{1_\mathbb{R}}}
\newcommand{\isatwoR}{\isamath{2_\mathbb{R}}}
\newcommand{\isaexists}{\isasymexists}
\newcommand{\isaforall}{\isasymforall}
\newcommand{\isaand}{\isasymand}
\newcommand{\isaN}{\isamath{\mathbb{N}}}
\newcommand{\isaR}{\isamath{\mathbb{R}}}
\newcommand{\isaRTop}{\isamath{\mathbb{R}}}
\newcommand{\isahalfR}{\isaoneR\ \isasub{/}{\mathbb{R}} \isatwoR}
\newcommand{\isapione}{\isamath{\pi_1}}
\newcommand{\isaRcal}{\isamath{\mathcal{R}}}
\newcommand{\isaScal}{\isamath{\mathcal{S}}}
\newcommand{\isamcomp}{\isamath{\circ_{\isasty{m}}}}

\allowdisplaybreaks[1]

\begin{document}

\title{Formalization of the fundamental group in untyped set theory
  using auto2}

\author{Bohua Zhan}

\institute{Massachusetts Institute of Technology}
\maketitle

\begin{abstract}
  We present a new framework for formalizing mathematics in untyped
  set theory using auto2. Using this framework, we formalize in
  Isabelle/FOL the entire chain of development from the axioms of set
  theory to the definition of the fundamental group for an arbitrary
  topological space. The auto2 prover is used as the sole automation
  tool, and enables succinct proof scripts throughout the project.
\end{abstract}

\section{Introduction}

Auto2, introduced by the author in \cite{auto2}, is a proof automation
tool for the proof assistant Isabelle. It is designed to be a
powerful, extensible prover that can consistently solve ``routine''
tasks encountered during a proof, thereby enabling a style of
formalization using succinct proof scripts written in a custom, purely
declarative language.

In this paper, we present an application of auto2 to formalization of
mathematics in untyped set theory \footnote{Code available at
  https://github.com/bzhan/auto2}. In particular, we discuss the
formalization in Isabelle/FOL of the entire chain of development from
the axioms of set theory to the definition of the fundamental group
for an arbitrary topological space. Along the way, we discuss several
improvements to auto2 as well as strategies of usage that allow us to
work effectively with untyped set theory.

The contribution of this paper is two-fold. First, we demonstrate that
the auto2 system is capable of independently supporting proof
developments on a relatively large scale. In the previous paper,
several case studies for auto2 were given in Isabelle/HOL. Each case
study is at most several hundred lines long, and the use of auto2 is
mixed with the use of other Isabelle tactics, as well as proof scripts
provided by Sledgehammer. In contrast, the example we present in this
paper is a unified development consisting of over 13,000 lines of
theory files and 3,500 lines of ML code (not including the core auto2
program). The auto2 prover is used exclusively starting from basic set
theory.

Second, we demonstrate one way to manage the additional complexity in
proofs that arise when working with untyped set theory. For a number
of reasons, untyped set theory is considered to be difficult to work
with. For example, everything is represented as sets, including
objects such as natural numbers that we usually do not think of as
sets. Moreover, statements of theorems tend to be longer in untyped
set theory than in typed theories, since assumptions that would
otherwise be included in type constraints must now be stated
explicitly. In this paper, we show that with appropriate definitions
of basic concepts and setup for automation, all these complexities can
be managed, without sacrificing the inherent flexibility of the logic.

We now give an outline for the rest of the paper. In Section
\ref{sec:structures}, we sketch our choice of definitions of basic
concepts in axiomatic set theory. In particular, we describe how to
use tuples to realize extensible records, and build up the hierarchy
of algebraic structures. In Section \ref{sec:auto2}, we review the
main ideas of the auto2 system, and describe several additional
features, as well as strategies of usage, that allow us to manage the
additional complexities of untyped set theory.

In Section \ref{sec:exampleselem}, we give two examples of proof
scripts using auto2, taken from the proofs of the Schroeder-Bernstein
theorem and a challenge problem in analysis from Lasse Rempe-Gillen.
In Section \ref{sec:fundamentalgroup}, we describe our main example,
the definition of the fundamental group, in detail. Given a
topological space $X$ and a base point $x$ on $X$, the fundamental
group $\pi_1(X,x)$ is defined on the quotient of the set of loops in
$X$ based at $x$, under the equivalence relation given by path
homotopy. Multiplication on $\pi_1(X,x)$ comes from joining two loops
end-to-end. Formalizing this definition requires reasoning about
algebraic and topological structures, equivalence relations, as well
as continuous functions on real numbers. We believe this is a
sufficiently challenging task with which to test the maturity of our
framework, although it has been achieved before in the Mizar
system. HOL Light and Isabelle/HOL also formalized the essential ideas
on path homotopy. We review these and other related works in Section
\ref{sec:relatedwork}, and conclude in Section \ref{sec:conclusion}.

\paragraph{Acknowledgements.}
The author would like to thank the anonymous referees for their
comments. This research is completed while the author is supported by
NSF Award No. 1400713.

\section{Basic constructions in set theory}
\label{sec:structures}

We now discuss our choice of definitions of basic concepts, starting
with the choice of logic. Our development is based on the FOL
(first-order logic) instantiation of Isabelle. The initial parts are
similar to those in Isabelle/ZF, and we refer to
\cite{paulson1,paulson2} for detailed explanations.

The only Isabelle types available are $i$ for sets, $o$ for
propositions (booleans), and function types formed from them. We call
objects with types other than $i$ and $o$ \emph{meta-functions}, to
distinguish them from functions defined within set theory (which have
type $i$). It is possible to define higher-order meta-functions in
FOL, and supply them with arguments in the form of lambda
expressions. Theorems can be quantified over variables with functional
type at the outermost level. These can be thought of as
theorem-schemas in a first-order theory. However, one can only
quantify over variables of type $i$ inside the statement of a theorem,
and the only equalities defined within FOL are those between types $i$
(notation $\cdot = \cdot$) and $o$ (notation $\cdot
\longleftrightarrow \cdot$). In practice, these restrictions mean that
any functions that we wish to consider as first-class objects must be
defined as set-theoretic functions.

\subsection{Axioms of set theory}\label{sec:axioms}

For uniformity of presentation, we start our development from FOL
rather than theories in Isabelle/ZF. However, the list of axioms we
use is mostly the same. The only main addition is the axiom of global
choice, which we use as an easier-to-apply version of the axiom of
choice. Note that as in Isabelle/ZF, several of the axioms introduce
new sets or meta-functions, and declare properties satisfied by
them. The exact list of axioms is as follows:

\begin{isabelle}
\ \ extension: \ \ "\isaforall z. z \isain\ x \isaiff\ z \isain\ y \isaImp\ x = y" \isanewline
\ \ empty\_set: \ \ "x \isasymnotin\ \isaempty" \isanewline
\ \ collect:\ \ \ \ \ "x \isain\ Collect(A,P) \isaiff\ (x \isain\ A \isaand\ P(x))" \isanewline
\ \ upair:\ \ \ \ \ \ \ "x \isain\ Upair(y,z) \isaiff\ (x = y \isasymor\ x = z)" \isanewline
\ \ union:\ \ \ \ \ \ \ "x \isain\ \isasymUnion C \isaiff\ (\isaexists A\isain C. x\isain A)" \isanewline
\ \ power:\ \ \ \ \ \ \ "x \isain\ Pow(S) \isaiff\ x \isasymsubseteq\ S" \isanewline
\ \ replacement: "\isaforall x\isain A. \isaforall y z. P(x,y) \isaand\ P(x,z) \isaimp\ y = z \isaImp \isanewline
\ \ \ \ \ \ \ \ \ \ \ \ \ \ \ \ \ b \isain\ Replace(A,P) \isaiff\ (\isaexists x\isain A. P(x,b))" \isanewline
\ \ foundation:\ \ "x \isasymnoteq\ \isaempty\ \isaImp\ \isaexists y\isain x. y \isasyminter\ x = \isaempty" \isanewline
\ \ infinity:\ \ \ \ "\isaempty\ \isain\ Inf \isaand\ (\isaforall y\isain Inf. succ(y) \isain\ Inf)" \isanewline
\ \ choice:\ \ \ \ \ \ "\isaexists x. x\isain S \isaImp\ Choice(S) \isain\ S"
\end{isabelle}

Next, we define several basic constructions in set theory. They are
summarized in the following table. See \cite{paulson1} for more
explanations.

\begin{tabular} {c|c}
  Notation & Definition \\ \hline
  \isa{THE x. P(x)} & \isa{\isasymUnion(Replace(\isaset{\isaempty}, \isasymlambda x y. P(y)))} \\
  \isa{\isaset{b(x). x\isain A}} & \isa{Replace(A, \isasymlambda x y. y = b(x))} \\
  \isa{SOME x\isain A. P(x)} & \isa{Choice(\isaset{x\isain A. P(x)})} \\
  \isa{\isapair{a,b}} & \isa{\isaset{\isaset{a}, \isaset{a,b}}} \\
  \isa{fst(p)} & \isa{THE a. \isaexists b. p = \isapair{a,b}} \\
  \isa{snd(p)} & \isa{THE b. \isaexists a. p = \isapair{a,b}} \\
  \isa{\isapair{\isamath{a_1,\dots,a_n}}} &
  \isa{\isapair{\isamath{a_1},\isapair{\isamath{a_2},\isapair{\isamath{\cdots,a_n}}}}} \\
  \isa{if P then a else b} & \isa{THE z. P \isaand\ z=a \isasymor\ \isasymnot P \isaand\ z=b} \\
  \isa{\isasymUnion a\isain I. X} & \isa{\isasymUnion \isaset{X(a). a\isain I}} \\
  \isa{A \isaprod\ B} & \isa{\isasymUnion x\isain A. \isasymUnion y\isain B. \isaset{\isapair{x,y}}}
\end{tabular}

\subsection{Extensible records as tuples}\label{sec:functions}

We now consider the problem of representing records. In our framework,
records are used to represent functions, algebraic and topological
structures, as well as morphisms between structures. It is often
advantageous for records of different types to share certain
fields. For example, groups and rings should share the multiplication
operator, rings and ordered rings should share both addition and
multiplication operators, and so on.

It is well-known that when formalizing mathematics using set theory,
records can be represented as tuples. To achieve sharing of fields,
the key idea is to assign each shared field a fixed position in the
tuple.

We begin with the example of functions. A function is a record
consisting of a source set (domain), a target set (codomain), and the
graph of the function. In particular, we consider two functions with
the same graph but different target sets to be different functions
(another structure called \emph{family} is used to represent functions
without specified target set). The three fields are assigned to the
first three positions in the tuple:

\begin{isabelle}
\isacommand{definition} "source(F) = fst(F)" \isanewline
\isacommand{definition} "target(F) = fst(snd(F))" \isanewline
\isacommand{definition} "graph(F) \ = fst(snd(snd(F)))"
\end{isabelle}

A function with source \isa{S}, target \isa{T}, and graph \isa{G} is
represented by the tuple \isa{\isapair{S,T,G,\isaempty}} (we append an
\isa{\isaempty} at the end so the definition of \isa{graph} works
properly). For \isa{G} to actually represent a function, it must
satisfy the conditions for a functional graph:

\begin{isabelle}
\isacommand{definition} func\_graphs :: "i \isaTo\ i \isaTo\ i" \isakeyword{where} \isanewline
\ \ "func\_graphs(X,Y) = \isalb G\isain Pow(X\isaprod Y). (\isamath{\forall}a\isain X. \isamath{\exists!}y. \isapair{a,y}\isain G)\isarb"
\end{isabelle}

The set of all functions from \isa{S} to \isa{T} (denoted \isa{S
  \isato\ T}) is then given by:

\begin{isabelle}
\isacommand{definition} function\_space :: "i \isaTo\ i \isaTo\ i" (\isakeyword{infixr} "\isato" 60) \isakeyword{where} \isanewline
\ \ "A \isato\ B = \isalb\isapair{A,B,G,\isaempty}. G\isain func\_graphs(A,B)\isarb"
\end{isabelle}

Functions can be created using the following constructor. Note this is
a higher-order meta-function. The argument \isa{b} can be supplied by
a lambda expression.
\begin{isabelle}
\isacommand{definition} Fun :: "[i, i, i \isaTo\ i] \isaTo\ i" where \isanewline
\ \ "Fun(A,B,b) = \isapair{A, B, \isaset{p\isain A\isaprod B. snd(p) = b(fst(p))}, \isaempty}"
\end{isabelle}

Evaluation of a function \isa{f} at \isa{x} (denoted \isa{f\isaeval
  x}) is then defined as:

\begin{isabelle}
\isacommand{definition} feval :: "i \isaTo\ i \isaTo\ i" (\isakeyword{infixl} "\isaeval" 90) \isakeyword{where} \isanewline
\ \ "f \isaeval\ x = (THE y. \isapair{x,y}\isain graph(f))"
\end{isabelle}

\subsection{Algebraic structures}

The second major use of records is to represent algebraic
structures. In our framework, we will define structures such as
groups, abelian groups, rings, and ordered rings. The carrier set of a
structure is assigned to the first position. The order relation,
additive data, and multiplicative data are assigned to the third,
fourth, and fifth position, respectively. This is expressed as
follows:

\begin{isabelle}
\isacommand{definition} "carrier(S) \ \ \ \ = fst(S)" \isanewline
\isacommand{definition} "order\_graph(S) = fst(snd(snd(S)))" \isanewline
\isacommand{definition} "zero(S) \ \ \ \ \ \ \ = fst(fst(snd(snd(snd(S)))))" \isanewline
\isacommand{definition} "plus\_fun(S) \ \ \ = snd(fst(snd(snd(snd(S)))))" \isanewline
\isacommand{definition} "one(S) \ \ \ \ \ \ \ \ = fst(fst(snd(snd(snd(snd(S))))))" \isanewline
\isacommand{definition} "times\_fun(S) \ \ = snd(fst(snd(snd(snd(snd(S))))))"
\end{isabelle}

Here \isa{order\_graph} is a subset of \isa{S\isaprod S}, and
\isa{plus\_fun}, \isa{times\_fun} are elements of \isa{S\isaprod
  S\isato S}. Hence, the operators $\le, +,$ and $*$ can be defined as
follows:

\begin{isabelle}
\isacommand{definition} "le(R,x,y) \isaiff\ \isapair{x,y}\isain order\_graph(R)" \isanewline
\isacommand{definition} "plus(R,x,y) = plus\_fun(R)\isaeval\isapair{x,y}" \isanewline
\isacommand{definition} "times(R,x,y) = times\_fun(R)\isaeval\isapair{x,y}" \isanewline
\end{isabelle}

These are abbreviated to \isa{x \isale{R} y}, \isa{x \isaplus{R} y},
and \isa{x \isatimes{R} y}, respectively (in both theory files and
throughout this paper, we use $*$ to denote multiplication in groups
and rings, and $\times$ to denote product on sets and other
structures). We also abbreviate \isa{x \isain\ carrier(S)} to \isa{x
  \isain. S}.

The constructor for group-like structures is as follows:
\begin{isabelle}
\isacommand{definition} Group :: "[i, i, i \isaTo\ i \isaTo\ i] \isaTo\ i" \isacommand{where} \isanewline
\ \ "Group(S,u,f) = \isapair{S,\isaempty,\isaempty,\isaempty,\isapair{u,\isasymlambda p\isain S\isaprod S. f(fst(p),snd(p))\isain S},\isaempty}"
\end{isabelle}

The following predicate asserts that a structure contains \emph{at
  least} the fields of a group-like structure, with the right
membership properties (\isa{\isaone{G}} abbreviates \isa{one(G)}):
\begin{isabelle}
\isacommand{definition} is\_group\_raw :: "i \isaTo\ o" \isacommand{where} \isanewline
\ \ "is\_group\_raw(G) \isaiff \isanewline
\ \ \ \ \ \isaone{G} \isain. G \isaand\ times\_fun(G) \isain\ carrier(G) \isaprod\ carrier(G) \isato\ carrier(G)
\end{isabelle}

To check whether such a structure is in fact a monoid / group, we use
the following predicates:
\begin{isabelle}
\isacommand{definition} is\_monoid :: "i \isaTo\ o" \isacommand{where} \isanewline
\ \ "is\_monoid(G) \isaiff\ is\_group\_raw(G) \isaand\ \isanewline
\ \ \ \ \ \ (\isaforall x\isain.G. \isaforall y\isain.G. \isaforall z\isain.G.
            (x \isatimes{G} y) \isatimes{G} z = x \isatimes{G} (y \isatimes{G} z)) \isaand \isanewline
\ \ \ \ \ \ (\isaforall x\isain.G. \isaone{G} \isatimes{G} x = x \isaand\ x \isatimes{G} \isaone{G} = x)"
\end{isabelle}

\begin{isabelle}
\isacommand{definition} units :: "i \isaTo\ i" \isacommand{where} \isanewline
\ \ "units(G) = \isaset{x \isain. G. (\isaexists y\isain.G. y \isatimes{G} x = \isaone{G} \isaand\ x \isatimes{G} y = \isaone{G})}"
\end{isabelle}

\begin{isabelle}
\isacommand{definition} is\_group :: "i \isaTo\ o" \isacommand{where} \isanewline
\ \ "is\_group(G) \isaiff\ is\_monoid(G) \isaand\ carrier(G) = units(G)"
\end{isabelle}

Note these definitions are meaningful on any structure that has
multiplicative data. Likewise, we can define a predicate
\isa{is\_abgroup} for abelian groups, that is meaningful for any
structure that has additive data. These can be combined with
distributive properties to define the predicate for a ring:

\begin{isabelle}
\isacommand{definition} is\_ring :: "i \isaTo\ o" \isacommand{where} \isanewline
\ \ "is\_ring(R) \isaiff\ (is\_ring\_raw(R) \isaand\ is\_abgroup(R) \isaand\ is\_monoid(R) \isaand \isanewline
\ \ \ \ \ \  is\_left\_distrib(R) \isaand\ is\_right\_distrib(R) \isaand\ \isazero{R} \isamath{\neq} \isaone{R})"
\end{isabelle}

Likewise, we can define the predicate for ordered rings, and
constructors for such structures. Structures are used to represent the
hierarchy of numbers: we let \isa{nat}, \isa{int}, \isa{rat}, and
\isa{real} denote the \emph{set} of natural numbers, integers, etc,
while \isamath{\mathbb{N}}, \isamath{\mathbb{Z}},
\isamath{\mathbb{Q}}, and \isamath{\mathbb{R}} denote the
corresponding structures. Hence, addition on natural numbers is
denoted by \isa{x \isaplusN\ y}, addition on real numbers by \isa{x
  \isaplusR\ y}, etc. We can also state and prove theorems such as
\isa{is\_ord\_field(\isamath{\mathbb{R}})}, which contains all proof
obligations for showing that the real numbers form an ordered field.

\subsection{Morphism between structures}

Finally, we discuss morphisms between structures. Morphisms can be
considered as an \emph{extension} of functions, with additional fields
specifying structures on the source and target sets. The two
additional fields are assigned to the fourth and fifth positions in
the tuple:

\begin{isabelle}
\isacommand{definition} "source\_str(F) = fst(snd(snd(snd(F))))" \isanewline
\isacommand{definition} "target\_str(F) = fst(snd(snd(snd(snd(F)))))"
\end{isabelle}

The constructor for a morphism is as follows (here \isa{S} and \isa{T}
are the source and target structures, while the source and target sets
are automatically derived):
\begin{isabelle}
\isacommand{definition} Mor :: "[i, i, i \isaTo\ i] \isaTo\ i" \isacommand{where} \isanewline
\ \ "Mor(S,T,b) = (let A = carrier(S) in let B = carrier(T) in \isanewline
\ \ \ \ \ \isapair{A, B, \isaset{p\isain A\isaprod B. snd(p) = b(fst(p))}, S, T, \isaempty})"
\end{isabelle}

The space of morphisms (denoted \isa{S \isamor\ T}) is given by:

\begin{isabelle}
\isacommand{definition} mor\_space :: "i \isaTo\ i \isaTo\ i" (\isakeyword{infix} "\isamor" 60) where \isanewline
\ \ "mor\_space(S,T) = (let A = carrier(S) in let B = carrier(T) in \isanewline
\ \ \ \ \ \isaset{{\isapair{A,B,G,S,T,\isaempty}. G\isain func\_graphs(A,B)}})"
\end{isabelle}

Note the notation \isa{f\isaeval x} for evaluation still works for
morphisms. Several other concepts defined in terms of evaluation, such
as image and inverse image, continue to be valid for morphisms as
well, as are lemmas about these concepts. However, operations that
construct new morphisms, such as inverse and composition, must be
redefined. We will use \isa{g \isamath{\circ} f} to denote the
composition of two functions, and \isa{g \isamcomp\ f} to denote the
composition of two morphisms.

Having morphisms store the source and target structures means we can
define properties such as homomorphism on groups as a predicate:

\begin{isabelle}
\isacommand{definition} is\_group\_hom :: "i \isaTo\ o" \isacommand{where} \isanewline
\ "is\_group\_hom(f) \isaiff\ (let S = source\_str(f) in let T = target\_str(f) in \isanewline
\ \ \ \ \ \ is\_morphism(f) \isaand\ is\_group(S) \isaand\ is\_group(T) \isaand\ \isanewline
\ \ \ \ \ \ (\isaforall x\isain.S. \isaforall y\isain.S. f\isaeval(x \isatimes{S} y) = f\isaeval x \isatimes{T} f\isaeval y))"
\end{isabelle}

The following lemma then states that the composition of two
homomorphisms is a homomorphism (this is proved automatically using
auto2):

\begin{isabelle}
lemma group\_hom\_compose: \isanewline
\ \ "is\_group\_hom(f) \isaImp\ is\_group\_hom(g) \isaImp\ \isanewline
\ \ \ target\_str(f) = source\_str(g) \isaImp\ is\_group\_hom(g \isamcomp\ f)"
\end{isabelle}

\section{Auto2 in untyped set theory}
\label{sec:auto2}

In this section, we describe several additional features of auto2, as
well as general strategies of using it to manage the complexities of
untyped set theory.

We begin with an overview of the auto2 system (see \cite{auto2} for
details). Auto2 is a theorem prover packaged as a tactic in
Isabelle. It works with a collection of rules of reasoning called
\emph{proof steps}. New proof steps can be added at any time within an
Isabelle theory. They can also be deleted at any time, although it is
rarely necessary to add and delete the same proof step more than
once. In general, when building an Isabelle theory, the user is
responsible for specifying, by adding proof steps, how to use the
results proved in that theory. In return, the user no longer needs to
worry about invoking these results by name in future developments.

The overall algorithm of auto2 is as follows. First, the statement to
be proved is converted into contradiction form, so the task is always
to derive a contradiction from a list of assumptions. During the
proof, auto2 maintains a list of \emph{items}, the two most common
types of which are propositions (that are derived from the
assumptions) and terms (that have appeared so far in the proof). Each
item resides in a \emph{box}, which can be thought of as a subcase of
the statement to be proved (the box corresponding to the original
statement is called the \emph{home box}). A proof step is a function
that takes as input one or two items, and outputs either new items,
new cases, or the action of shadowing one of the input items, or
resolving a box by proving a contradiction in that box.

The main loop of the algorithm repeatly applies the current collection
of proof steps and adds any new items and cases in a best-first-search
manner, until some proof step derives a contradiction in the home
box. In addition to the list of items, auto2 also maintains several
tables. The most important of which is the \emph{rewrite table}, which
keeps track of the list of currently known equalities (not containing
arbitrary variables), and maintains the congruence closure of these
equalities. There are two other tables: the property table and the
well-form table, which we will discuss later in this section.

There are two broad categories of proof steps, which we call the
\emph{standard} and \emph{special} proof steps in this paper. A
standard proof step applies an existing theorem in a specific
direction. It matches the input items to one or two patterns in the
statement of the theorem, and applies the theorem to derive a new
proposition. Here the matching is up to rewriting (\emph{E-matching})
using the rewrite table. A special proof step can have more complex
behavior, and is usually written as an ML function. The vast majority
of proof steps in our example are standard, although special proof
steps also play an important role.

The auto2 prover is not intended to be complete. For example, it may
intentionally apply a theorem in only one of several possible
directions, in order to narrow the search space. For more difficult
theorems, auto2 provides a custom language of proof scripts, allowing
the user to specify intermediate steps of the proof. Generally, when
proving a result using auto2, the user will first try to prove it
without any scripts, and in case of failure, successively add
intermediate steps, perhaps by referring to an informal proof of the
result. In case of failure, auto2 will indicate the first intermediate
step that it is unable to prove, as well as what it is able to derive
in the course of proving that step. We will show examples of proof
scripts in Section \ref{sec:exampleselem}.

The current version of auto2 can be set up to work with different
logics in Isabelle. It contains a core program, for reasoning about
predicate logic and equality, that is parametrized over the list of
constants and theorems for the target logic. In particular, auto2 is
now set up and tested to work with both HOL and FOL in Isabelle.

\subsection{Encapsulation of definitions}\label{sec:definition}

One commonly cited problem with untyped set theory is that every
object is a set, including those that are not usually considered as
sets. Common examples of the latter include ordered pairs, natural
numbers, functions, etc. In informal treatments of mathematics, these
definitions are only used to establish some basic properties of the
objects concerned. Once these properties are proved, the definitions
are never used again.

In formal developments, when automation is used to produce large parts
of the proof, one potential problem is that the automation may
needlessly expand the original definitions of objects, rather than
focusing on their basic properties. This increases the search space
and obscures the essential ideas of the proof. Using the ability to
delete proof steps in auto2, this problem can be avoided entirely. For
any definition that we wish to drop in the end, we use the following
three-step procedure:

\begin{enumerate}
\item The definition is stated and added to auto2 as rewrite rules.
\item Basic properties of the object being defined are stated and
  proved. These properties are added to auto2 as appropriate proof
  steps.
\item The rewrite rules for the original definition are deleted.
\end{enumerate}

For example, after the definitions concerning the representation of
functions as tuples in Section \ref{sec:functions}, we prove the
following lemmas, and add them as appropriate proof steps (as
indicated by the attributes in brackets):

\begin{isabelle}
\isacommand{lemma} lambda\_is\_function [backward]: \isanewline
\ \ "\isaforall x\isain A. f(x)\isain B \isaImp\ Fun(A,B,f) \isain\ A \isato\ B"
\end{isabelle}

\begin{isabelle}
\isacommand{lemma} beta [rewrite]: \isanewline
\ \ "F = Fun(A,B,f) \isaImp\ x \isain\ source(F) \isaImp\ is\_function(F) \isaImp\ F\isaeval x = f(x)"
\end{isabelle}

\begin{isabelle}
\isacommand{lemma} feval\_in\_range [typing]: \isanewline
\ \ "is\_function(f) \isaImp\ x \isain\ source(f) \isaImp\ f\isaeval x \isain\ target(f)"
\end{isabelle}

After proving these (and a few more) lemmas, the rewriting rules for
the definitions of \isa{Fun}, \isa{function\_space}, \isa{feval}, etc,
are removed. Note that all lemmas above are independent of the
representation of functions as tuples. Hence, this representation is
effectively hidden from the point of view of the prover. Some of the
original definitions may be temporarily re-added in rare instances
(for example when defining the concept of morphisms).

\subsection{Property and well-form tables}\label{sec:tables}

In this section, we discuss two additional tables maintained by auto2
during a proof. The property table is already present in the version
introduced in \cite{auto2}, but not discussed in that paper. The
well-form table is new.

The main motivation for both tables is that for many theorems,
especially those stated in an untyped logic, some of its assumptions
can be considered as ``side conditions''. To give a basic example,
consider the following lemma:

\begin{isabelle}
\isacommand{lemma} unit\_l\_cancel: \isanewline
\ \ "is\_monoid(G) \isaImp\ y \isain. G \isaImp\ z \isain. G \isaImp\ x \isatimes{G} y = x \isatimes{G} z \isaImp\ \isanewline
\ \ \ x \isain\ units(G) \isaImp\ y = z"
\end{isabelle}

In this lemma, the last two assumptions are the ``main'' assumptions,
while the first three are side conditions asserting that the variables
in the main assumptions are well-behaved in some sense. In
Isabelle/HOL, these side conditions may be folded into type or
type-class constraints.

We consider two kinds of side conditions. The first kind, like the
first assumption above, checks that one of the variables in the main
assumptions satisfy a certain predicate. In Isabelle/HOL, these may
correspond to type-class constraints. In auto2, we call these
\emph{property assumptions}. More precisely, given any predicate (in
FOL this means constant of type \isa{i \isaTo\ o}), we can register it
as a property. The \emph{property table} records the list of
properties satisfied by each term that has appeared so far in the
proof. Properties propagate through equalities: if \isa{P(a)} is in
the property table, and \isa{a = b} is known from the rewrite table,
then \isa{P(b)} is automatically added to the property table. The user
can also add theorems of certain forms as further propagation rules
for the property table (we omit the details here).

The second kind of side conditions assert that certain terms occuring
in the main assumptions are \emph{well-formed}. We use the terminology
of well-formedness to capture a familiar feature of mathematical
language: that an expression may make implicit assumptions about its
subterms. These conditions can be in the form of type constraints. For
example, the expression \isa{a \isaplus{R} b} implicitly assumes that
\isa{a} and \isa{b} are elements in the carrier set of
\isa{R}. However, this concept is much more general. Some examples of
well-formedness conditions are summarized in the following table:

\begin{tabular} {c|c}
  Term & Conditions \\ \hline
  \isa{\isasymInter A} & \isa{A \isamath{\neq} \isaempty} \\
  \isa{f \isaeval\ x} & \isa{x \isain\ source(f)} \\
  \isa{g \isamath{\circ} f} & \isa{target(f) = source(g)} \\
  \isa{g \isamcomp\ f} & \isa{target\_str(f) = source\_str(g)} \\
  \isa{a \isaplus{R} b} & \isa{a \isain. R, b \isain. R} \\
  \isa{inv(R,a)} & \isa{a \isain\ units(R)} \\
  \isa{a \isamath{/_{\isasty{R}}} b} & \isa{a \isain. R, b \isain\ units(R)} \\
  \isa{subgroup(G,H)} & \isa{is\_subgroup\_set(G,H)} \\
  \isa{quotient\_group(G,H)} & \isa{is\_normal\_subgroup\_set(G,H)}
\end{tabular}

In general, given any meta-function \isa{f}, any propositional
expression in terms of the arguments of \isa{f} can be registered as a
well-formedness condition of \isa{f}. In particular, well-formedness
conditions are not necessarily properties. For example, the condition
\isa{a \isain. R} for \isa{a \isaplus{R} b} involves two variables and
hence is not a property. The \emph{well-form table} records, for every
term encountered so far in the proof, the list of its well-formedness
conditions that are satisfied. Whenever a new fact is added, auto2
checks against every known term to see whether it verifies a
well-formedness condition of that term.

The property and well-form tables are used in similar ways in standard
proof steps. After the proof step matches one or two patterns in the
``main'' assumptions or conclusion of the theorem that it applies, it
checks for the side conditions in the two tables, and proceed to apply
the theorem only if all side conditions are found. Of course, this
requires proof steps to be re-applied if new properties or
well-formedness conditions of a term becomes known.

\subsection{Well-formed conversions}\label{sec:conversion}

Algebraic simplification is an important part of any automatic
prover. For every kind of algebraic structure, e.g. monoids, groups,
abelian groups, and rings, there is a concept of normal form of an
expression, and two terms can be equated if they have the same normal
form. In untyped set theory, such computation of normal forms is
complicated by the fact that the relevant rewriting rules have extra
assumptions. For example, the rule for associativity of addition is:

\begin{isabelle}
\ \ is\_abgroup(R) \isaImp\ x \isain. R \isaImp\ y \isain. R \isaImp\ z \isain. R \isaImp\ \isanewline
\ \ \ \ \ \ x \isaplus{R} (y \isaplus{R} z) = (x \isaplus{R} y) \isaplus{R} z
\end{isabelle}

The first assumption can be verified at the beginning of the
normalization process. The remaining assumptions, however, are more
cumbersome. In particular, they may require membership status of terms
that arise only during the normalization. For example, when
normalizing the term \isa{a\isaplus{R}(b\isaplus{R}(c\isaplus{R}d))},
we may first rewrite it to
\isa{a\isaplus{R}((b\isaplus{R}c)\isaplus{R}d)}. The next step,
however, requires \isa{b\isaplus{R}c \isain. R}, where
\isa{b\isaplus{R}c} does not occur initially and may not have occured
so far in the proof. In typed theories, this poses no problem, since
\isa{b\isamath{+}c} will be automatically given the same type as
\isa{b} and \isa{c} when the term is created.

In untyped set theory, such membership information must be kept track
of and derived when necessary. The concept of well-formed terms
provides a natural framework for doing this. Before performing
algebraic normalization on a term, we first check for all relevant
well-formedness conditions. If all conditions are present, we produce
a data structure (of type \isa{wfterm} in Isabelle/ML) containing the
certified term as well as theorems asserting well-formedness
conditions. A theorem is called a \emph{well-formed rewrite rule} if
its main conclusion is an equality, each of its assumptions is a
well-formedness condition for terms on the left side of the equality,
and it has additional conclusions that verify all well-formedness
conditions for terms on the right side of the equality that are not
already present in the assumptions. For example, the associativity
rule stated above is not yet a well-formed rewrite rule: there is no
justification for \isa{x\isaplus{R}y \isain. R}, which is a
well-formedness condition for the term
\isa{(x\isaplus{R}y)\isaplus{R}z} on the right side of the
equality. The full well-formed rewrite rule is:

\begin{isabelle}
\ \ is\_abgroup(R) \isaImp\ x \isain. R \isaImp\ y \isain. R \isaImp\ z \isain. R \isaImp\ \isanewline
\ \ \ \ \ \ x \isaplus{R} (y \isaplus{R} z) = (x \isaplus{R} y) \isaplus{R} z \isaand\ x \isaplus{R} y \isain. R
\end{isabelle}

Given a well-formed rewrite rule, we can produce a \emph{well-formed
  conversion} that acts on \isa{wfterm} objects, in a way similar to
how equalities produce regular conversions that act on \isa{cterm}
objects in Isabelle/ML. Like regular conversions, well-formed
conversions can be composed in various ways, and full normalization
procedures can be written using the language of well-formed
conversions. These normalization procedures in turn form the basis of
several special proof steps. We give two examples:

\begin{itemize}
\item Given two terms $s$ and $t$ that are non-atomic with respect to
  operations in \isa{R}, where \isa{R} is a monoid (group / abelian
  group / ring), normalize $s$ and $t$ using the rules for \isa{R}. If
  the normalizations are equal, output $s = t$.
\item Given two propositions \isa{a \isale{R} b} and \isa{\isasymnot
    (c \isale{R} d)}, where \isa{R} is an ordered ring. Compare the
  normalizations of \isa{b \isaminus{R} a} and \isa{d \isaminus{R}
    c}. If they are equal, output a contradiction.
\end{itemize}

These proof steps, when combined with proof scripts provided by the
user, allow algebraic manipulations to be performed rapidly. They
replace the handling of associative-commutative functions for HOL
discussed in \cite{auto2}.

\subsection{Discussion}\label{sec:discussion}

We conclude this section with a discussion of our overall approach to
untyped set theory, and compare it with other approaches. One feature
of our approach is that we do not seek to re-institute a concept of
types in our framework, but simply replace type constraints with set
membership conditions (or predicates, for constraints that cannot be
described by a set). The aim is to fully preserve the flexibility of
set-membership as compared to types. Empirically, most of the extra
assumptions that arise in the statement of theorems can be taken care
of by classifying them as properties or well-formedness
conditions. Our approach can be contrasted with that taken by Mizar,
which defines a concept of soft types \cite{Mizar-type1} within the
core of the system.

Every framework for formalizing modern mathematics need a way to deal
with structures. In Mizar, structures are defined in the core of the
system as partial functions on selectors
\cite{Mizar-struct,Mizar-struct2}. In both Isabelle/HOL and
IsarMathLib's treatement of abstract algebra, structures are realized
with extensive use of locales. For Coq, one notable approach is the
use of Canonical Structures \cite{coq-canonical} in the formalization
of the Odd Order Theorem. We chose a relatively simple scheme of
realizing structures as tuples, which is sufficient for the present
purposes. Representing them as partial functions on selectors, as in
Mizar, is more complicated but may be beneficial in the long run.

Finally, we emphasize that we do not make any modification to
Isabelle/FOL in our development. The concept of well-formed terms, for
example, is meaningful only to the automation. The whole of auto2's
design, including the ability for users to add new proof steps,
follows the LCF architecture. To have confidence in the proofs, one
only need to trust the existing Isabelle system, the ten axioms stated
in Section \ref{sec:axioms}, and the definitions involved in the
statement of the results.

\section{Examples of proof scripts}
\label{sec:exampleselem}

Using the techniques in the above two sections, we formalized enough
mathematics in Isabelle/FOL to be able to define the fundamental
group. In addition to work directly used for that purpose, we also
formalized several interesting results on the side. These include the
well-ordering theorem and Zorn's lemma, the first isomorphism theorem
for groups, and the intermediate value theorem. Two more examples will
be presented in the remainder of this section, to demonstrate the
level of succinctness of proof scripts that can be achieved.

Throughout our work, we referred to various sources including both
mathematical texts and other formalizations. We list these sources
here:

\begin{itemize}
\item Axioms of set theory and basic operations on sets, construction
  of natural numbers using least fixed points: from Isabelle/ZF
  \cite{paulson1,paulson2}.
\item Equivalence and order relations, arbitrary products on sets,
  well-ordering theorem and Zorn's lemma: from Bourbaki's \emph{Theory
    of Sets} \cite{bourbaki}.
\item Group theory and the construction of real numbers using Cauchy
  sequences: from my previous case studies \cite{auto2}, which in turn
  is based on corresponding theories in the Isabelle/HOL library.
\item Point-set topology and construction of the fundamental group:
  from \emph{Topology} by Munkres \cite{munkres}.
\end{itemize}

\subsection{Schroeder-Bernstein Theorem}

For our first example, we present the proof of the Schroeder-Bernstein
theorem. See \cite{paulson2} for a presentation of the same proof in
Isabelle/ZF. The bijection is constructed by gluing together two
functions. Auto2 is able to prove automatically that under certain
conditions, the gluing is a bijection (lemma
\isa{glue\_function2\_bij}). For the Schroeder-Bernstein theorem, a
proof script (provided by the user) is needed. This is given
immediately after the statement of the theorem.

\begin{isabelle}
\isacommand{definition} glue\_function2 :: "i \isaTo\ i \isaTo\ i" \isacommand{where} \isanewline
\ \ "glue\_function2(f,g) = Fun(source(f)\isasymunion source(g), target(f)\isasymunion target(g), \isanewline
\ \ \ \ \ \isasymlambda x. if x \isain\ source(f) then f\isaeval x else g\isaeval x)''
\end{isabelle}

\begin{isabelle}
\isacommand{lemma} glue\_function2\_bij [backward]: \isanewline
\ \ "f \isain\ A \isasymcong\ B \isaImp\ g \isain\ C \isasymcong\ D \isaImp\ A \isasyminter\ C = \isaempty\ \isaImp\ B \isasyminter\ D = \isaempty\ \isaImp \isanewline
\ \ \ glue\_function2(f,g) \isain\ (A \isasymunion\ C) \isasymcong\ (B \isasymunion\ D)"
\end{isabelle}

\begin{isabelle}
\isacommand{theorem} schroeder\_bernstein: \isanewline
\ \ "injective(f) \isaImp\ injective(g) \isaImp\ f \isain\ X \isato\ Y \isaImp\ g \isain\ Y \isato\ X \isaImp \isanewline
\ \ \ equipotent(X,Y)" \isanewline
\ \ LET "X\_A = lfp(X, \isasymlambda W. X -- g\isaimage (Y -- f\isaimage W))" THEN \isanewline
\ \ LET "X\_B = X -- X\_A, Y\_A = f\isaimage X\_A, Y\_B = Y -- Y\_A" THEN \isanewline
\ \ HAVE "X -- g\isaimage Y\_B = X\_A" THEN \isanewline
\ \ HAVE "g\isaimage Y\_B = X\_B" THEN \isanewline
\ \ LET "f' = func\_restrict\_image(func\_restrict(f,X\_A))" THEN \isanewline
\ \ LET "g' = func\_restrict\_image(func\_restrict(g,Y\_B))" THEN \isanewline
\ \ HAVE "glue\_function2(f', inverse(g')) \isain\ (X\_A \isasymunion\ X\_B) \isasymcong\ (Y\_A \isasymunion\ Y\_B)"
\end{isabelle}

\subsection{Rempe-Gillen's challenge}

For our second example, we present our solution to a challenge problem
proposed by Lasse Rempe-Gillen in a mailing list discussion
\footnote{http://www.cs.nyu.edu/pipermail/fom/2014-October/018243.html}. See
\cite{hammering} for proofs of the same result in several other
systems. The statement to be proved is:

\begin{lemma}
Let $f$ be a continuous real-valued function on the real line, such
that $f(x) > x$ for all $x$. Let $x_0$ be a real number, and define
the sequence $x_n$ recursively by $x_{n+1} := f(x_n)$. Then $x_n$
diverges to infinity.
\end{lemma}

Our solution is as follows. We make use of several previously proved
results: any bounded increasing sequence in $\mathbb{R}$ converges
(line 2), a continuous function \isa{f} maps a sequence converging to
\isa{x} to a sequence converging to \isa{f\isaeval x} (line 4), and
finally that the limit of a sequence in $\mathbb{R}$ is unique.
\begin{isabelle}
\isacommand{lemma} rempe\_gillen\_challenge: \isanewline
\ \ "real\_fun(f) \isaImp\ continuous(f) \isaImp\ incr\_arg\_fun(f) \isaImp\ x0 \isain. \isaR\ \isaImp \isanewline
\ \ \ S = Seq(\isaR, \isasymlambda n. nfold(f,n,x0)) \isaImp\ \isasymnot upper\_bounded(S)" \isanewline
\ \ HAVE "seq\_incr(S)" WITH HAVE "\isaforall n\isain.\isaN. S\isaeval (n \isaplusR\ 1) \isageR\ S\isaeval n" THEN \isanewline
\ \ CHOOSE "x, converges\_to(S,x)" THEN \isanewline
\ \ LET "T = Seq(\isaR, \isasymlambda n. f\isaeval (S\isaeval n))" THEN \isanewline
\ \ HAVE "converges\_to(T,f\isaeval x)" THEN \isanewline
\ \ HAVE "converges\_to(T,x)" WITH ( \isanewline
\ \ \ \ HAVE "\isaforall r\isagtR \isazeroR. \isaexists k\isain.\isaN. \isaforall n\isageN k. \isasymbar T\isaeval n \isaminusR x\isasymbar\isasub{}{\mathbb{R}} \isaltR\ r" WITH ( \isanewline
\ \ \ \ \ \ CHOOSE "k \isain. \isaN, \isaforall n\isageN k. \isasymbar S\isaeval n \isaminusR\ x\isasymbar\isasub{}{\mathbb{R}} \isaltR\ r" THEN \isanewline
\ \ \ \ \ \ HAVE "\isaforall n\isageN k. \isasymbar T\isaeval n \isaminusR\ x\isasymbar\isasub{}{\mathbb{R}} \isaltR\ r" WITH HAVE "T\isaeval n = S\isaeval (n \isaplusN\ 1)"))
\end{isabelle}

\section{Construction of the fundamental group}
\label{sec:fundamentalgroup}

In this section, we describe our construction of the fundamental
group. We will focus on stating the definitions and main results
without proof, to demonstrate the expressiveness of untyped set theory
under our framework. The entire formalization including proofs is 864
lines long.

Let \isa{I} be the interval \isa{[0,1]}, equipped with the subspace
topology from the topology on $\mathbb{R}$. Given two continuous maps
\isa{f} and \isa{g} from \isa{S} to \isa{T}, a \emph{homotopy} between
\isa{f} and \isa{g} is a continuous map from the product topology on
\isa{S \isaprod\ I} to \isa{T} that restricts to \isa{f} and \isa{g}
at \isa{S \isaprod\ \isaset{0}} and \isa{S \isaprod\ \isaset{1}},
respectively:

\begin{isabelle}
\isacommand{definition} is\_homotopy :: "[i, i, i] \isaTo\ o" \isacommand{where} \isanewline
\ \ "is\_homotopy(f,g,F) \isaiff\ \isanewline
\ \ \ \ \ (let S = source\_str(f) in let T = target\_str(f) in \isanewline
\ \ \ \ \ \ continuous(f) \isaand\ continuous(g) \isaand\ \isanewline
\ \ \ \ \ \ S = source\_str(g) \isaand\ T = target\_str(g) \isaand\ F \isain\ S \isasub{\times}{\isasty{T}} I \isasub{\isamor}{\isasty{T}} T \isaand \isanewline
\ \ \ \ \ \ (\isaforall x\isain.S. F\isaeval\isapair{x,\isazeroR} = f\isaeval x \isaand\ F\isaeval\isapair{x,\isaoneR} = g\isaeval x))"
\end{isabelle}

A \emph{path} is a continuous function from the interval. A homotopy
between two paths is a \emph{path homotopy} if it remains constant on
\isa{\isaset{0} \isaprod\ I} and \isa{\isaset{1} \isaprod\ I}:

\begin{isabelle}
\isacommand{definition} is\_path :: "i \isaTo\ o" \isacommand{where} \isanewline
\ \ "is\_path(f) \isaiff\ (f \isain\ I \isasub{\isamor}{\isasty{T}} target\_str(f))"
\end{isabelle}

\begin{isabelle}
\isacommand{definition} is\_path\_homotopy :: "[i, i, i] \isaTo\ o" \isacommand{where} \isanewline
\ \ "is\_path\_homotopy(f,g,F) \isaiff\ \isanewline
\ \ \ \ (is\_path(f) \isaand\ is\_path(g) \isaand\ is\_homotopy(f,g,F) \isaand \isanewline
\ \ \ \ \ (\isaforall t\isain.I. F\isaeval\isapair{\isazeroR,t} = f\isaeval(\isazeroR) \isaand\ F\isaeval\isapair{\isaoneR,t} = f\isaeval(\isaoneR)))"
\end{isabelle}

Two paths are \emph{path-homotopic} if there exists a path homotopy
between them. This is an equivalence relation on paths.

\begin{isabelle}
\isacommand{definition} path\_homotopic :: "i \isaTo\ i \isaTo\ o" \isacommand{where} \isanewline
\ \ "path\_homotopic(f,g) \isaiff\ (\isaexists F. is\_path\_homotopy(f,g,F))"
\end{isabelle}

The path product is defined by gluing two morphisms. It is continuous
by the pasting lemma:

\begin{isabelle}
\isacommand{definition} I1 = subspace(\isaRTop, closed\_interval(\isaR,\isazeroR,\isahalfR)) \isanewline
\isacommand{definition} I2 = subspace(\isaRTop, closed\_interval(\isaR,\isahalfR,\isaoneR)) \isanewline
\isacommand{definition} interval\_lower = Mor(I1,I,\isasymlambda t. \isatwoR\ \isatimesR\ t) \isanewline
\isacommand{definition} interval\_upper = Mor(I2,I,\isasymlambda t. \isatwoR\ \isatimesR\ t\ \isaminusR\ \isaoneR)
\end{isabelle}

\begin{isabelle}
\isacommand{definition} path\_product :: "i \isaTo\ i \isaTo\ i"  (infixl "\isasymstar" 70) where \isanewline
\ \ "f \isasymstar\ g = glue\_morphism(I, f \isamcomp\ interval\_lower, g \isamcomp\ interval\_upper)"
\end{isabelle}

The loop space is a set of loops on \isa{X}. Path homotopy gives an
equivalence relation on the loop space, and we define
\isa{loop\_classes} to be the quotient set:

\begin{isabelle}
\isacommand{definition} loop\_space :: "i \isaTo\ i \isaTo\ i" \isacommand{where} \isanewline
\ \ "loop\_space(X,x) = \isaset{f \isain\ I \isasub{\isamor}{\isasty{T}} X. f\isaeval(\isazeroR) = x \isaand\ f\isaeval(\isaoneR) = x}"
\end{isabelle}

\begin{isabelle}
\isacommand{definition} loop\_space\_rel :: "i \isaTo\ i \isaTo\ i" \isacommand{where} \isanewline
\ \ "loop\_space\_rel(X,x) = Equiv(loop\_space(X,x), \isasymlambda f g. path\_homotopic(f,g))"
\end{isabelle}

\begin{isabelle}
\isacommand{definition} loop\_classes :: "i \isaTo\ i \isaTo\ i" \isacommand{where} \isanewline
\ \ "loop\_classes(X,x) = loop\_space(X,x) // loop\_space\_rel(X,x)"
\end{isabelle}

Finally, the fundamental group is defined as:

\begin{isabelle}
\isacommand{definition} fundamental\_group :: "i \isaTo\ i \isaTo\ i" ("\isapione") where \isanewline
\ \ "\isapione(X,x) = (let \isaRcal\ = loop\_space\_rel(X,x) in \isanewline
\ \ \ \ \ Group(loop\_classes(X,x), equiv\_class(\isaRcal,const\_mor(I,X,x)), \isanewline
\ \ \ \ \ \ \ \ \ \ \ \isasymlambda f g. equiv\_class(\isaRcal,rep(\isaRcal,f) \isasymstar\ rep(\isaRcal,g))))"
\end{isabelle}

To show that the fundamental group is actually a group, we need to
show that the path product respects the equivalence relation given by
path homotopy, and is associative up to equivalence (along with
properties about inverse and identity). The end result is:

\begin{isabelle}
\isacommand{lemma} fundamental\_group\_is\_group: \isanewline
\ \ "is\_top\_space(X) \isaImp\ x \isain. X \isaImp\ is\_group(\isapione(X,x))"
\end{isabelle}

An important property of the fundamental group is that a continuous
function between topological spaces induces a homomorphism between
their fundamental groups. This is defined as follows:

\begin{isabelle}
\isacommand{definition} induced\_mor :: "i \isaTo\ i \isaTo\ i" \isacommand{where} \isanewline
\ \ "induced\_mor(k,x) = \isanewline
\ \ \ \ (let X = source\_str(k) in let Y = target\_str(k) in \isanewline
\ \ \ \ \ let \isaRcal\ = loop\_space\_rel(X,x) in let \isaScal\ = loop\_space\_rel(Y,k\isaeval x) in \isanewline
\ \ \ \ \ Mor(\isapione(X,x), \isapione(Y,k\isaeval x), \isasymlambda f. equiv\_class(\isaScal, k \isamcomp\ rep(\isaRcal,f))))"
\end{isabelle}

The induced map is a homomorphism satisfying functorial properties:

\begin{isabelle}
\isacommand{lemma} induced\_mor\_is\_homomorphism: \isanewline
\ \ "continuous(k) \isaImp\ X = source\_str(k) \isaImp\ Y = target\_str(k) \isaImp \isanewline
\ \ \ x \isain\ source(k) \isaImp\ induced\_mor(k,x) \isain\ \isapione(X,x) \isasub{\isamor}{\isasty{G}} \isapione(Y,k\isaeval x)"
\end{isabelle}

\begin{isabelle}
\isacommand{lemma} induced\_mor\_id: \isanewline
\ \ "is\_top\_space(X) \isaImp\ x \isain. X \isaImp \isanewline
\ \ \ induced\_mor(id\_mor(X),x) = id\_mor(\isapione(X,x))"
\end{isabelle}

\begin{isabelle}
\isacommand{lemma} induced\_mor\_comp: \isanewline
\ \ "continuous(k) \isaImp\ continuous(h) \isaImp \isanewline
\ \ \ target\_str(k) = source\_str(h) \isaImp\ x \isain\ source(k) \isaImp \isanewline
\ \ \ induced\_mor(h \isamcomp\ k, x) = induced\_mor(h, k\isaeval x) \isamcomp\ induced\_mor(k, x)"
\end{isabelle}

\section{Related work} \label{sec:relatedwork}

In Isabelle, the main library for formalized mathematics using FOL is
Isabelle/ZF. The basics of Isabelle/ZF is described in
\cite{paulson1,paulson2}. We also point to \cite{paulson1} for a
review of older work on set theory from automated deduction and
artificial intelligence communities. Outside the official library,
IsarMathLib \cite{isarmathlib} is a more recent project based on
Isabelle/ZF. It formalized more results in abstract algebra and
point-set topology, and also constructed the real numbers. The initial
parts of our development closedly parallels that in Isabelle/ZF, but
we go further in several directions including constructing the number
system. The primary difference between our work and IsarMathLib is
that we use auto2 for proofs, and develop our own system for handling
structures, so that we do not make use of Isabelle tactics, Isar, or
locales.

Outside Isabelle, the major formalization projects using set theory
include Metamath \cite{metamath} and Mizar \cite{Mizar}, both of which
have extensive mathematical libraries. There are some recent efforts
to reproduce the Mizar environment in HOL-type systems
\cite{Mizar-HOL2,Mizar-HOL1}. While there are some similarities
between our framework and Mizar's, we do not aim for an exact
reproduction. In particular, we maintain the typical style of stating
definitions and theorems in Isabelle. More comparisons between our
approach and Mizar are discussed in Section \ref{sec:discussion}.

Mizar formalized not just the definition of the fundamental group
\cite{fundamental-group-mizar}, but several of its properties,
including the computation of the fundamental group of the
circle. There is also a formalization of path homotopy in HOL Light
which is then ported to Isabelle/HOL. This is used for the proof of
the Brouwer fixed-point theorem and the Cauchy integral theorem,
although the fundamental group itself does not appear to be
constructed.

In homotopy type theory, one can work with fundamental groups (and
higher-homotopy groups) using synthetic definitions. This has led to
formalizations of results about homotopy groups that are well beyond
what can be achieved today using standard definitions (see
\cite{hott-homotopy-group} for a more recent example). We emphasize
that our definition of the fundamental group, as with Mizar's, follows
the standard one in set theory.

\section{Conclusion} \label{sec:conclusion}

We applied the auto2 prover to the formalization of mathematics using
untyped set theory. Starting from the axioms of set theory, we
formalized the definition of the fundamental group, as well as many
other results in set theory, group theory, point-set topology, and
real analysis. The entire development contains over 13,000 lines of
theory files and 3,500 lines of ML code, taking the author about 5
months to complete. On a laptop with two 2.0GHz cores, it can be
compiled in about 24 minutes. Through this work, we demonstrated the
ability of auto2 to scale to relatively large projects. We also hope
this result can bring renewed interest to formalizing mathematics in
untyped set theory in Isabelle.


\begin{thebibliography}{10}

\bibitem{hammering} Blanchette, J. C., Kaliszyk, C., Paulson, L. C.,
  Urban, J.: Hammering towards QED. Journal of Formalized Reasoning
  9(1), pp.101--148, 2016.

\bibitem{bourbaki} Bourbaki, N.: Theory of Sets. Springer, 2000

\bibitem{hott-homotopy-group} Brunerie, G.: On the homotopy groups of
  spheres in homotopy type
  theory. Ph.D. Thesis. https://arxiv.org/abs/1606.05916

\bibitem{Mizar} Grabowski, A., Kornilowicz, A., Naumowicz, A.: Mizar
  in a nutshell. J. Formaliz. Reason. Spec. Issue: User Tutor. I 3(2),
  153--245, 2010.

\bibitem{isarmathlib} IsarMathLib: http://www.nongnu.org/isarmathlib/

\bibitem{Mizar-HOL2} Kaliszyk C., Pak, K., and Urban, J.: Towards a
  Mizar environment for Isabelle: foundations and language. In
  Proceedings of the 5th ACM SIGPLAN Conference on Certified Programs
  and Proofs (CPP 2016). ACM, New York, NY, USA, 58--65, 2016.

\bibitem{fundamental-group-mizar} Kornilowicz, A., Shidama, Y.,
  Grabowski, A.. The Fundamental Group, Formalized Mathematics 12(3),
  pages 261--268, 2004.

\bibitem{Mizar-HOL1} Kuncar, O.. Reconstruction of the Mizar type
  system in the HOL Light system. In J. Pavlu and J. Safrankova,
  editors, WDS Proceedings of Contributed Papers: Part I --
  Mathematics and Computer Sciences, pages 7--12. Matfyzpress, 2010.

\bibitem{coq-canonical} Mahboubi, A., Tassi, E.: Canonical Structures
  for the Working Coq User. In: Blazy S., Paulin-Mohring C., Pichardie
  D. (eds) ITP 2013. LNCS, vol 7998. Springer, Berlin, Heidelberg,
  pp. 19--34, 2013.

\bibitem{metamath} Megill, N.D.: Metamath: A computer language for
  pure mathematics. http://us.metamath.org/downloads/metamath.pdf

\bibitem{munkres} Munkres, J.R.: Topology. Prentice Hall, 2000

\bibitem{paulson1} Paulson, L.C.: Set theory for verification: I. From
  foundations to functions. In: Journal of Automated Reasoning,
  11(3):353--389, 1993.

\bibitem{paulson2} Paulson, L.C.: Set theory for verification:
  II. Induction and recursion. In: Journal of Automated Reasoning,
  15(2): 167--215, 1995.

\bibitem{Mizar-struct} Trybulec, A.: Some Features of the Mizar
  Language. ESPRIT Workshop, 1993.

\bibitem{Mizar-struct2} Lee, G., Rudnicki, P.: Alternative Aggregates
  in Mizar. In: Kauers M., Kerber M., Miner R., Windsteiger W. (eds)
  Towards Mechanized Mathematical Assistants. LNCS, vol
  4573. Springer, Berlin, Heidelberg, 2007.

\bibitem{Mizar-type1} Wiedijk, F.: Mizar's Soft Type System. In:
  Schneider K., Brandt J. (eds) Theorem Proving in Higher Order
  Logics. LNCS, vol 4732. Springer, Berlin, Heidelberg, 2007.

\bibitem{auto2} Zhan, B.: AUTO2: a saturation-based heuristic prover
  for higher-order logic. In: J.C. Blanchette and S. Merz (Eds.): ITP
  2016, LNCS 9807, pp. 441--456, 2016.

\end{thebibliography}
\end{document}